\documentclass{article}
 \usepackage{latexsym}
 \usepackage{amsfonts}

\newtheorem{theoreme}{Theorem }[section]
\newtheorem{proposition}[theoreme]{Proposition}
\newtheorem{lemma}[theoreme]{Lemma}
\newtheorem{definition}[theoreme]{Definition}

\newcommand{\beq}{\begin{equation}}
\newcommand{\eeq}{\end{equation}}

\newcounter{smallarabics}
\newenvironment{arabicenumerate}
{\begin{list}{{\normalfont\textrm{(\arabic{smallarabics})}}}
  {\usecounter{smallarabics}\setlength{\itemindent}{0cm}
   \setlength{\leftmargin}{5ex}\setlength{\labelwidth}{4ex}
   \setlength{\topsep}{0.75\parsep}\setlength{\partopsep}{0ex}
   \setlength{\itemsep}{0ex}}}
{\end{list}}

\newcounter{smallroman}

\def\bel{\begin{lemma}}
\def\eel{\end{lemma}}
\def\bet{\begin{theoreme}}
\def\eet{\end{theoreme}}
\def\bed{\begin{definition}}
\def\eed{\end{definition}}
\def\bep{\begin{proposition}}
\def\eep{\end{proposition}}
\def\ben{\begin{arabicenumerate}}
\def\een{\end{arabicenumerate}}

\def\rr{{\mathbb R}}
\def\zz{{\mathbb Z}}
\def\cc{{\mathbb C}}

\def\i{{\rm i}}

\def\e{{\rm e}}
\def\d{{\rm d}}

\def\tg{{\rm tg}}
\def\tgh{{\rm tgh}}
\def\ctgh{{\rm ctgh}}

\def\cG{{\cal G}}

\def\P{{\cal P}}

\def\F{{\cal F}}
\def\cF{{\cal F}}
\def\cA{{\cal A}}

\def\12{\frac{1}{2}}

\def\p{ \partial}

\def\trho{{\tilde \rho}}

\begin{document}
\title{
Exactly solvable Schr\"odinger operators
}
 \author{J. Derezi\'{n}ski
\\
Department of Mathematical Methods in Physics,  
\\ Warsaw University,\\ Ho\.{z}a 74, 00-682 Warszawa, Poland,\\
email jan.derezinski@fuw.edu.pl\\
\\
M. Wrochna
\\
RTG ``Mathematical Structures in Modern Quantum Physics'',
\\ Institute of Mathematics, University of G\"ottingen,\\ Bunsenstr. 3-5, D-37073 G\"ottingen, Germany
\\ email wrochna@uni-math.gwdg.de}
\maketitle 
\tableofcontents

\begin{abstract}
We systematically describe and classify
 1-dimensional Schr\"odinger equations that can be solved in terms of 
hypergeometric type functions. Beside the
 well-known families, we explicitly describe 2 new
classes of exactly solvable Schr\"odinger equations that
 can be reduced to the Hermite equation.
\end{abstract}

\section{Introduction}

Exactly solvable 1-dimensional Schr\"odinger equations play an important role
in quantum physics. The best known are the harmonic oscillator and the radial
equation for the hydrogen atom, which are covered in every course of quantum
mechanics. A number of other examples were discovered in the 30's of the last
century. They include (known also under other names)  trigonometric and hyperbolic  P\"{o}schl-Teller 
potentials \cite{pt}, the   Scarf potential \cite{scarf},
the Eckart potential
 \cite{eckart}, the
 Manning-Rosen potential \cite{mrosen} and the  Rosen-Morse potential.
 All these examples can be reduced to
 the hypergeometric equation
(see eg.
\cite{ww}). One should also mention the Morse potential
 that leads to the
 confluent equation.
Problems involving these potentials are often used in classes on quantum
mechanics, 
see eg. the well-known problem book of
 Fl\"ugge \cite{fluegge}.  A number of techniques have been developed to study
 their properties, such as the factorization method  \cite{hi} and the
closely related method of the  superpotential (see for instance \cite{cks} and \cite{cotfas}). 

In later years J.~N.~Ginocchio \cite{ginocchio}
  discovered that these examples can be
generalized to a larger class of potentials equivalent to the hypergeometric or
confluent equation. Later, this class was extended by G.~A.~Natanzon
\cite{natanzon} and further generalized by R.~Milson \cite{milson}. These classes, besides the hitherto known potentials, are not
very practical in applications, since they are not given by explicit
expressions. 

There exist even more general classes of potentials that can be called exactly
solvable, found by A.~Khare and U.~Sukhatme \cite{ks}. They are, however,
expressed in exotic special functions.

The literature on exactly solvable Schr\"odinger  equations is very large. The
subject is in fact very useful for applications, especially in quantum
physics. We also believe that it is quite beautiful, which in the existing
literature is perhaps not so easy to see.

 In our paper we would like to 
 systematically  describe basic classes of exactly
solvable
Schr\"odinger  equations.
Most of the material of our paper is scattered in the literature, notably in
\cite{natanzon}, \cite{milson} and \cite{cks} (see also \cite{bose}). 
Our treatment is, however,
somewhat
 more systematic than what we could
 find in the literature. For instance, we
distinguish between the {\em complex classification} and the {\em real
  classification}. 
 It also seems that for
the first time we explicitly describe 2 new classes of exactly solvable
Schr\"odinger equations, which can be reduced to the Hermite equation (see
Subsects \ref{specI} and \ref{specII}).

By a  {\em (stationary 1-dimensional)
Schr\"odinger  equation} we will mean an  equation  of the form
\beq \left(-\p_r^2+V(r)-E\right)\phi(r)=0.\label{schro3}\eeq
This equation can be interpreted as the eigenvalue problem for the operator
\beq H:=-\p_r^2+V(r).\label{schro}\eeq
An operator of the form (\ref{schro})
will be  called a {\em (1-dimensional) 
Schr\"odinger operator}. $V(r)$ will be called a {\em potential} and the
parameter $E$ an {\em energy}.

(\ref{schro}) can be interpreted as an operator in a  number of ways. If $V$
is a holomorphic function of the complex variable on some open $\Omega\subset
\cc$,
 then $r$ can be interpreted as a complex variable and
(\ref{schro}) can be viewed as an operator on holomorphic functions. In this
 case it is natural to allow  $E$ to be a
 complex parameter. The corresponding eigenvalue equation
 (\ref{schro3}) 
will be then called the  {\em Schr\"odinger equation of the complex
   variable}. Let us note that,  {\em complex affine
transformations} preserve the class of Schr\"odinger
 equations  of the  complex variable.
(By a complex affine
transformations we mean $r\mapsto ar +b$, where
$a\neq0,$ $b$ are complex constants).

One can also interpret $r$ as a real variable in some open 
  $I\subset\rr$. The operator (\ref{schro}) is then viewed as acting on
functions on $I$.
Understood in this way, (\ref{schro3}) will be called
 the {\em Schr\"odinger equation of the real variable}. 
Of special interest is then 
the case of {\em real potential}. Clearly,
 {\em real affine
transformations} preserve the class of real Schr\"odinger
 equations  of the  real variable.

Our paper is organized as follows.
First,
we will briefly 
dicuss some general facts related to 2nd order linear differential
equations. In particular, we will describe basic ingredients of the so-called
{\em Bose-Natanzon method}, which permits to obtain a class of Schr\"odinger
equations equivalent to a chosen linear equation. Note
 that it applies to more general situations than those described in the
 literature (\cite{natanzon}, \cite{milson}). We will formulate a criterion
 that determines when the Bose-Natanzon method can be used. 

Next, we will focus on \textit{hypergeometric type equations}.
Recall that equations of the form
\beq\left(\sigma(z)\p_z^2+\tau(z)\p_z+
\eta\right) f(z)=0,
\label{req}\eeq
where
 $\sigma(z),\ \tau(z)$ are polynomials with
\[{\deg}\sigma\leq 2,\ 
{\deg}\tau\leq 1,\]
and $\eta$ is a number,
are sometimes called
 {\em  hypergeometric type
  equations}
(see A.F.~Nikiforov and V.B.~Uvarov \cite{nu}). Solutions of (\ref{req}) are
called {\em hypergeometric type functions}. Hypergeometric type functions are very
well understood and include
the hypergeomeric function, the confluent function, Hermite, Laguerre and
Jacobi polynomials, etc. Traditionally, one  classifies
 hypergeometric type equations into
several distinct classes invariant
under complex affine 
transformations.
In each class one chooses a simple representative,
 to which the whole class can be
reduced. Such representatives are the hypergeometric equation,
the confluent equation, etc. 

In our analysis, we will  concentrate on 3 specific
equations: 
\ben 
\item the {\em hypergeometric equation}
\beq \left(z(1-z)\p_z^2+(c-(a+b+1)z)\p_z-ab\right)f(z)=0;\label{hyp1}\eeq
\item the {\em rescaled confluent equation}
\beq\left(z\p_z^2+(c-\gamma z)\p_z-a\right)f(z)=0.\label{hyp2}\eeq
\item
the {\em translated harmonic  oscillator}
\beq\left(-\partial_z^2+\theta^2 z^2+\rho
z+\lambda\right)f(z)=0.\label{harmo}\eeq 
\een

Note that 
all   hypergeometric type equations with a nonzero $\sigma$
are related with one of the above
equations by elementary operations
\begin{itemize}\item multiplication of $f(z)$ by a function,
multiplication 
of the equation by a function,\item a change of variables not depending on
the parameters $a,b,c$ (respectively $a,c,\gamma$ or $\theta, \rho,\lambda$).  
\end{itemize}

\ben\item If $\sigma$ is 2nd order and has 2 distinct roots, an affine
transformation reduces  (\ref{req}) to (\ref{hyp1}).
\item If $\sigma$ is $2$nd order and has only one root, then a transformation
involving  $z\mapsto z^{-1}$ 
reduces (\ref{req}) to
  (\ref{hyp2}) with $\gamma\neq0$. 
\item If  $\sigma$ is 1st order, then an affine transformation reduces 
(\ref{req}) to
  (\ref{hyp2}). 
\item
If $\sigma$ is  0th order and nonzero, then (\ref{req}) is equivalent to
 (\ref{harmo}) with $\theta\neq0$.
\een

Obviously, (\ref{hyp1}) and (\ref{hyp2}) are themselves equations of the
hypergeometric type. (\ref{harmo}) is not, but can be reduced to a
hypergeometric type equation. This reduction is  depends on whether
 $\theta\neq0$ or $\theta=0$. The former case leads to a hypergeometric type
equation with a constant $\sigma$, as mentioned above.
 In the case $\theta=0$, (\ref{harmo}) is the Airy
equation, which  can be reduced to a special case of the hypergeometric
type equation by a transformation involving $z\mapsto z^3$.

Equations  (\ref{hyp1}), (\ref{hyp2}) and (\ref{harmo}) are very well
understood
and have well known
solutions. 
We will describe classes of solvable potentials reducible to one of the
equations (\ref{hyp1}), (\ref{hyp2}) or (\ref{harmo}). We will describe both
the 
complex and the real classification of such potentials.

It is natural to
consider first
a classification of exactly solvable Schr\"odinger
equations of the complex variable. 
Note that all solvable  potentials that we consider are holomorphic
on the complex plane, apart from some isolated
singularities. 
 Obviously, we can always use a complex
affine 
transformation to put the equation in a convenient form. One can also move 
 a
complex 
constant from the potential to the energy.

Consider
a family of exactly solvable holomorphic potentials.
Suppose that for a selected subfamily
 of parameters the potential  is real if restricted to an open,
possibly infinite interval $I\subset\cc$. By elementary properties of
holomorphic functions,
if we  extend the interval $I$, the  potential is still real until we hit a
singularity. 
A real affine transformation can be used to put the
 equation in a convenient form. In particular, we can always assume that $I$
 is a subset of the real line. We can also move a real constant from the
 potential to the energy.

 The above discussion  motivates the following definition:
 We will say that an open interval
$I\subset\rr$ is a
{\em natural real domain} for a certain family of potentials if
it
 ends  either at $-\infty$, $+\infty$ or at a singularity of the
 potential, and no singularities lie inside $I$. When describing the real
 classification, we will always specify a natural real domain of the
 potential.

Clearly, if $I$ is a natural real domain, then 
the operator $H=-\partial_r^2+V(r)$ is
 hermitian on $C_{\rm c}^\infty(I)$.
 It is then natural to ask about 
 self-adjoint extensions
 of $H$. We will not discuss this question here. We
 plan to 
 consider it in later papers.

In Equations  (\ref{hyp1}), (\ref{hyp2}) and (\ref{harmo}) we have 3
arbitrary parameters.
Therefore, in 
all the solvable cases that we will describe,
 the potential  depends on 2 parameters, since
 the third parameter is
 responsible for the energy $E$. 



{\bf Acknowledgement} The research of J.~D.
was supported in part by the grant N N201 270135 of
the Polish Ministry of Science and Higher Education. 

\section{Second order homogeneous differential equations}

The main object of our paper are ordinary homogeneous 2nd order
linear differential
equations, that is equations of the form
\beq \left(a(r)\partial^2_r+b(r)\partial_r+c(r)\right)\phi(r)=0.\label{equo}\eeq
 It will be convenient to treat (\ref{equo}) as the problem of
finding the kernel of the operator
\beq\cA(r,\partial_r):=
a(r)\partial^2_r+b(r)\partial_r+c(r).\label{equo2}\eeq
We will then say that the equation (\ref{equo}) is given by the operator (\ref{equo2}).

We will treat $r$ either as a complex or a real variable.
In the complex case we will usually assume that the coefficients are
analytic.

In this section we describe some general facts related to the theory of
equations of the form (\ref{equo}) and their reduction to the Schr\"odinger
equation. 

\subsection{The Bose invariant
of a  second order differential equation}

Suppose that
we are given a second order differential equation (\ref{equo}).
Then we can always eliminate the first order term as follows.
We divide from the left by $a(r)$, set
\[h(r):=\exp\left(\int^{r}\frac{b(t)}{2a(t)}\d t\right).\]
We check that
\[h(r)\cA(r,\partial_r)h(r)^{-1}=\partial_r^2+I(r),\]
where, using the notation $a' := \frac{\d}{\d r} a(r)$,
\beq I=\frac{4ac-2ab'+2ba'-b^2}{4a^2}.\label{bose}\eeq
Clearly, $\check\phi(r):=h(r)\phi(r)$ solves the equation
\beq
\left(\partial^2_r+I(r)\right)\check\phi(r)=0.\label{bose1}\eeq

(\ref{bose1}) will be called the canonical form of the equation  (\ref{equo}).
(\ref{bose}) will be called the {\em Bose invariant} of (\ref{equo})
 \cite{bose}. (The name ``Bose invariant''
is used eg. by Milson \cite{milson}, however the object itself was clearly
known before Bose).

It is easy to check that for any functions $f,g$ the equation
\beq
f(r)\cA(r,\partial_r)g(r)\phi(r)=0
\label{equo1}\eeq  
has the same Bose invariant as (\ref{equo}). Conversely, if an equation has
the same Bose invariant as (\ref{equo}), it equals (\ref{equo1}) for some $f,g$.

\subsection{Schwarz derivative}

 The {\em Schwarz derivative} 
of  $r\mapsto r(y)$ appears naturally in the context of 2nd order ordinary
 differential equations and was known to XIX century mathematicians. It was
 mentioned already by Kummer in 1836 \cite{kummer}. The universally adopted
 name Schwarz derivative  was introduced by A.~Cayley, who derived its basic
 properties in 1880 \cite{cayley}. For more historical comments together with
 numerous applications we refer the reader to a book by Osgood \cite{osgood}.

 The Schwarz derivative is
defined as
\begin{equation}\label{eq:rr_schwartz}
\left\{r,y\right\}=\left({\frac{r''(y)}{r'(y)}}\right)'-\frac{1}{2}\left({\frac{r''(y)}{r'(y)}}\right)^2.
\end{equation}

Here are some basic properties of the Schwarz derivative:
\bep \begin{enumerate}\item Let $y\mapsto s(y)$, $s\mapsto r(s)$ be two functions. Then
\begin{equation}\label{eq:schwartz_chain}
\left\{r,y\right\}=\left\{r,s\right\}s'(y)^2+\left\{s,y\right\}.
\end{equation}
\item Let $r\mapsto y(r)$ be the inverse of $y\mapsto r(y)$. Then
\begin{equation}
\left\{y,r\right\}=-[r'(y)]^{-2}\left\{r,y\right\}.
\end{equation}
\item 
$\left\{r,y\right\}\equiv 0 \Leftrightarrow r(y)=\frac{ay+b}{cy+d}$, $ad-bc\neq 0$
\end{enumerate}
\eep

We will need the following fact:
\bep Suppose that $z'=z^p(1-z)^q$. Then
\beq
\{z,r\}=\frac12z^{2p-2}(1-z)^{2q-2}
\left((p^2-2p)(1-z)^2+(q^2-2q)z^2-2pqz(1-z)\right).\label{schwarz1}\eeq
\eep
  
\subsection{The Liouville transformation}

Consider a 2nd order equation in the canonical form, that is
\begin{equation}\label{eq:rr_postac_kanoniczna2}
(\partial^2_r+I(r))\phi(r)=0.
\end{equation}
Let us make the transformation
 $r =r(y)$ in this equation.
We obtain
\begin{equation}\label{eq:rr_zamiana_zmiennych}
\frac{1}{[r'(y)]^2}\partial^2_r\phi(r(y))-\frac{r''(y)}{[r'(y)]^3}\partial_r
\phi(r(y))+I(r(y)) 
\phi(r(y))=0
\end{equation}
The resulting equation we transform again to its canonical form:
\begin{equation}\label{eq:rr_postac_kanoniczna3}
(\partial^2_y+J(y)){\psi}(y)=0,
\end{equation}
where
\begin{equation}\label{eq:rr_definicja_psi}
\psi(y)=\frac{1}{[r'(y)]^2}\phi(r(y)),
\end{equation}
\begin{equation}\label{eq:rr_definicja_J}
J(y)=[r'(y)]^2I(r(y))+\frac{1}{2}\left\{r,y\right\},
\end{equation}
The above procedure is called the {\em Liouville transformation of
 (\ref{eq:rr_postac_kanoniczna2}) by the change of variables}
 $r = r(y)$ \cite{liouville}. One can check using (\ref{eq:rr_zamiana_zmiennych}), that the composition of two Liouville transformations, first by the change of variables $r=r(y)$, then $y=y(z)$, is the Liouville transformation by the change of variables $r=r(y(z))$.  

\subsection{Natanzon's problem}

Let $\cA(a_1,\dots,a_k;z,\partial_z)$ be a 2nd order differential operator
depending on $k$ parameters, such that its Bose invariant can be written as 
\beq\label{eq:assumptionI}
I(b_1,\dots,b_n;z)=b_1 I_1(z) + \dots + b_n I_n(z) 
\eeq
for some (at least one) $n\leq k$ linearly independent functions
$I_1,\dots,I_n$ (independent of $a_1,\dots,a_k$), 
 and some $n$ numbers $b_i(a_1,\dots,a_k)$. 
Then one can solve the following problem (which we will call
\textit{Natanzon's problem} for $\cA(z,\partial_z)$): 

\ \\
{\em Find all potentials $V(r)$ such that the 1-dimensional stationary
  Schr\"odinger equation
\beq
\left(-\partial_r^2+V(r)-E\right)\phi(E,r)=0\label{schro1}\eeq
 can be transformed to the equation given by $\cA(a_1,\dots,a_k;z,\partial_z)$ for some $a_1(E),\dots,a_k(E)$. We allow the following
operations:
\ben\item multiplication of both sides of the equation by some $f(E,r)$;
\item substitution of $\check\phi(E,r):=g(E,r)\phi(E,r)$ for some $g(E,r)$;
\item change of coordinates $r\mapsto z$  independent of $E$.
\een}
\ \\
This is the problem solved by Natanzon in the case of the hypergeometric
equation and by Milson in the general case of hypergeometric type
equations. In the following, we recall the construction they used, in a
slightly generalized form. 

Let us consider an arbitrary equation of the form (\ref{schro1}). Obviously,
its Bose invariant  equals $E-V(r)$. Clearly, the  transformations (1)
and (2) 
allow us to transform $\cA(a_1,\dots,a_k;z,\partial_z)$ to its canonical form 
\beq -\partial_z^2-I(b_1,\dots,b_n;z).\eeq
Thus   (\ref{schro1}) can be
 transformed to $\cA(a_1,\dots,a_k;z,\partial_z)$ if
\beq
I(b_1,\dots,b_n;z)=\left(r'(z)\right)^2(E-V\bigl(r(z)\bigr)
+\frac12\{r,z\}, \label{khk}\eeq
that is, if the two Bose invariants are related  by a Liouville
transformation. Using (\ref{eq:assumptionI}) we rewrite (\ref{khk})  as 
\beq
b_1(E) I_1(z) + \dots + b_n(E) I_n(z)=\left(r'(z)\right)^2(E-V\bigl(r(z)\bigr)
+\frac12\{r,z\}.
\eeq  
By assumption, $b_i$ depend on $E$ and $I_i(z)$ are independent of
$E$. Thus, the dependence of $b_i$ on $E$ is linear. Therefore, by
transforming linearly $(b_1,\dots,b_n)$ into $(\tilde b_1,\dots, \tilde b_{n-1}, E)$, we can assume that
\[I(b_1,\dots,b_n;z)=\tilde b_1 \tilde I_1(z)+\dots+\tilde b_{n-1} \tilde I_{n-1}(z)+ E \tilde
I_n(z),\]
for some functions $\tilde I_i(z)$, which are linear combinations of
$I_i(z)$.  Then we can write
\beq
E \tilde I_n(z)+\tilde b_1 \tilde I_1(z)+\dots+\tilde b_{n-1} \tilde
I_{n-1}(z)
=\left(r'(z)\right)^2(E-V\bigl(r(z)\bigr)
+\frac12\{r,z\}.\eeq 
Therefore, Natanzon's problem is solved by the following pair of equations
\begin{eqnarray}\nonumber
V\bigl(\tilde b_1,\dots,\tilde b_{n-1},r(z)\bigr)&:=&\left(r'(z)\right)^{-2}
\left(\tilde b_1 \tilde I_1(z)+\dots+\tilde b_{n-1} \tilde I_{n-1}(z)
-\frac12\{r,z\}\right),\\
\left(r'(z)\right)^2&=&\tilde I_n(z).\label{nata1}
\end{eqnarray}
We can rewrite (\ref{nata1}) in an equivalent way as
\begin{eqnarray}\nonumber
V\bigl(\tilde b_1,\dots,\tilde b_{n-1},r\bigr)&:=&\left(z'(r)\right)^{2}
\left(\tilde b_1 \tilde I_1(z(r))+\dots+\tilde b_{n-1} \tilde I_{n-1}(z(r))\right)
+\frac12\{z,r\},\\
\left(z'(r)\right)^{-2}&=&\tilde I_n(z(r)).\label{nata2}
\end{eqnarray}
If desired, we can renormalize $E$ by substracting a constant.

Note that in practical applications the value of the above method depends on
whether it is possible to invert the relation $r\mapsto z$ expressing it in
terms of standard functions. In the next sections, we will describe such
solutions of the Natanzon problem for the three classes of equations described
in the introduction. 


\section{Schr\"odinger operators reducible to the hypergeometric equation}

\subsection{Hypergeometric equation}

The {\em hypergeometric equation} is given by the operator
\beq \F(a,b;c;z,\p_z):=
z(1-z)\p_z^2+(c-(a+b+1)z)\p_z-ab,\eeq
where $a,b,c$ are arbitrary complex parameters.
Its Bose invariant equals
\begin{eqnarray*}
{}_2I_1(a,b;c;z)
&=&\frac{(1-a^2-b^2-2ab)z^2+2(-2ab+ac+bc-c)z+2c-c^2}{4z^2(1-z)^2}
.
\end{eqnarray*}

It satisfies condition (\ref{eq:assumptionI}) for $n=3$. The functions $I_i(z)$, $i=1,2,3$ can be taken to be
 $\frac{1}{4z^2(1-z)}$
 $\frac{1}{4z(1-z)^2}$
and $\frac{1}{4z(1-z)}$. It is natural to demand that $\left(r'(z)\right)^2=\tilde
I_3(z)$ is a function of the 
form $z^{2p}(1-z)^{2q}$. It is easy to see that this gives the following
possibilities for $\tilde I_3(z)$:
\begin{eqnarray}\nonumber
&& (1)\ \frac{1}{4z(1-z)},\ \ (2)\ 
\frac{1}{4z^2(1-z)},\ \ (3)\ \frac{1}{4z(1-z)^2},
\\
&&(4)\ \frac{1}{4z^2(1-z)^2},\ \ (5) \frac{1}{4z^2},\ \ 
(6) \frac{1}{4(1-z)^2}.\label{cases}
\end{eqnarray}
We will see that each of these ansatzes leads to exactly solvable potentials
considered in the literature. Note also that the formula for the Schwarz
derivative (\ref{schwarz1}) will be handy.

\subsection{The Riemann equation}

It is well known that it is useful to consider the hypergeometric equation as a special case of the so-called
{\em Riemann equation}. The Riemann equation is defined as the class of equations
on the Riemann sphere $\bar\cc$ having  3 regular singular points.
The following theorem summarizes the basic theory of these equations:

\bet \ben \item Suppose that we are given a 2nd order differential equation
 on the Riemann sphere
having
 3 singular points $z_1,z_2,z_3$, all of them regular singular points
with the following indices
\[\begin{array}{l}
z_1:\ \ \ \ \rho_1,\trho_1,\\
z_2:\ \ \ \ \rho_2,\trho_2,\\
z_3:\ \ \ \ \rho_3,\trho_3.
\end{array}\]
Then the following condition is satisfied:
\beq\rho_1+\trho_1+\rho_2+\trho_2+\rho_3+\trho_3=1.\label{ff3}\eeq
 Such an equation, normalized to have coefficient $1$ at the 2nd derivative,
 is always equal to
\beq
\P\left(\begin{array}{cccc}
z_1&z_2&z_3&\\
\rho_1&\rho_2&\rho_3&z,\p_z\\
\trho_1&\trho_2&\trho_3&\end{array}\right)\phi(z)=0,\label{ff1}\eeq
where
\[\begin{array}{r}\P\left(\begin{array}{cccc}
z_1&z_2&z_3&\\
\rho_1&\rho_2&\rho_3&z,\p_z\\
\trho_1&\trho_2&\trho_3&\end{array}\right):=
\p_z^2-\left(\frac{\rho_1+\trho_1-1}{z-z_1}+
\frac{\rho_2+\trho_2-1}{z-z_2}+
\frac{\rho_3+\trho_3-1}{z-z_3}\right)\p_z\\[6mm]
+\frac{\rho_1\trho_1(z_1-z_2)(z_1-z_3)}{(z-z_1)^2(z-z_2)(z-z_3)}
+\frac{\rho_2\trho_2(z_2-z_3)(z_2-z_1)}{(z-z_2)^2(z-z_3)(z-z_1)}
+\frac{\rho_3\trho_3(z_3-z_1)(z_3-z_2)}{(z-z_3)^2(z-z_1)(z-z_2)}
.\end{array}\]
\item Let $z\mapsto w =h(z)=\frac{az+b}{cz+d}$.
(Transformations of this form are called  homographies or M\"obius
transformations).  We can always assume that
$ad-bc=1$. Then
\[\P\left(\begin{array}{cccc}
h(z_1)&h(z_2)&h(z_3)&\\
\rho_1&\rho_2&\rho_3&w,\p_w\\
\trho_1&\trho_2&\trho_3&\end{array}\right)=
(cz+d)^4\P\left(\begin{array}{cccc}
z_1&z_2&z_3&\\
\rho_1&\rho_2&\rho_3&z,\p_z\\
\trho_1&\trho_2&\trho_3&\end{array}\right),
\]
\item
\[\begin{array}{l}
(z-z_1)^{-\lambda}(z-z_2)^\lambda\P\left(\begin{array}{cccc}
z_1&z_2&z_3&\\
\rho_1&\rho_2&\rho_3&z,\p_z\\
\trho_1&\trho_2&\trho_3&\end{array}\right)
(z-z_1)^{\lambda}(z-z_2)^{-\lambda}
\\[5mm]=\P\left(\begin{array}{cccc}
z_1&z_2&z_3&\\
\rho_1-\lambda&\rho_2+\lambda&\rho_3&z,\p_z\\
\trho_1-\lambda&\trho_2+\lambda&\trho_3&\end{array}\right).
\end{array}\]
\een\label{riema}\eet

Clearly, in all above formulas one of $z_i$ can equal $\infty$, with an
obvious meaning of various expressions. For convenience we 
give the expression for the Riemann operator with $z_3=\infty$:
\begin{eqnarray}\nonumber
&&\P\left(\begin{array}{cccc}
z_1&z_2&\infty&\\
\rho_1&\rho_2&\rho_3&z,\p_z\\
\trho_1&\trho_2&\trho_3&\end{array}\right)
\\
&=&
\p_z^2-\left(\frac{\rho_1+\trho_1-1}{z-z_1}+
\frac{\rho_2+\trho_2-1}{z-z_2}
\right)\p_z\nonumber \\
&&+\frac{\rho_1\trho_1(z_1-z_2)}{(z-z_1)^2(z-z_2)}
+\frac{\rho_2\trho_2(z_2-z_1)}{(z-z_2)^2(z-z_1)}
+\frac{\rho_3\trho_3}{(z-z_1)(z-z_2)}
\label{ff2}\end{eqnarray}

The  hypergeometric equation is a special case of the Riemann equation, since
\[\begin{array}{rl}
\F(a,b;c;z,\p_z)&=z(1-z)\P\left(\begin{array}{cccc}
0&1&\infty&\\
0&0&a&z,\p_z\\
1-c&c-a-b&b&\end{array}\right)
\\[7mm]&=z(1-z)\p_z^2+(c-(a+b+1)z)\p_z-ab.
\end{array}\]
Every Riemann equation can be brought to the form of the hypergeometric
equation by applying (1) and (2) of Theorem
\ref{riema}.

From Theorem \ref{riema} we also see that
  symmetries of the hypergeometric equation are better visible
 if we replace
 $a,b,c$ with $\alpha,\beta ,\mu$:
\[\begin{array}{rrr}\label{newnot}
\alpha=c-1&\ \ \beta =a+b-c,&\ \ \mu=a-b;\\
a=\frac{1+\alpha+\beta -\mu}{2},&\ \ b=\frac{1+\alpha+\beta
  +\mu}{2},&\ \ c=1+\alpha. 
\end{array}\]
In fact, the new parameters coincide with the differences of the 
indices of the points $0,1,\infty$:
\[\alpha=\rho_1-\trho_1,\ \beta=\rho_2-\trho_2,\ \mu=\rho_3-\trho_3.\]
In the parameters $\alpha,\beta,\mu$, the Bose invariant of the hypergeometric
equation has a more symmetric
form:
\begin{eqnarray*}
{}_2I_1(\alpha,\beta,\mu;z)
&=&\frac{(1-\alpha^2)(1-z)+(1-\beta^2)z+(\mu^2-1)(1-z)z}{4z^2(1-z)^2}.
\end{eqnarray*}
We can summarize the relation between the hypergeometric equation and its
canonical form by
\begin{eqnarray}
&&-z^{\frac{\alpha}{2}+\frac{1}{2}-1}(1-z)^{\frac{\beta}{2}+\frac{1}{2}-1}
\F\left({\textstyle\frac{\alpha+\beta+\mu+1}{2}},
{\textstyle\frac{\alpha+\beta-\mu+1}{2}}
;1+\alpha;z,\p_z\right)
z^{-\frac{\alpha}{2}-\frac{1}{2}}(1-z)^{-\frac{\beta}{2}-\frac{1}{2}}
\nonumber\\[2ex]
&=&-\P\left(\begin{array}{cccc}
0&1&\infty&\\
\frac{\alpha}{2}+\frac{1}{2}&\frac{\beta}{2}+\frac{1}{2}&\frac{\mu}{2}-\frac12&z,\p_z\\
-\frac{\alpha}{2}+\frac{1}{2}&-\frac{\beta}{2}+\frac{1}{2}
&-\frac{\mu}{2}-\frac12 
&\end{array}\right)\nonumber\\[4ex]
&=&-\partial_z^2
+\left(\alpha^2-1\right)\frac{1}{4z^2(1-z)}
+\left(\beta^2-1\right)\frac{1}{4z(1-z)^2}
-\left(\mu^2-1\right)\frac{1}{4z(1-z)}.\label{riem7}
\end{eqnarray}
It will be natural to introduce the parameters
\beq\kappa:=\frac12(\alpha^2-\beta^2),\ \ 
\delta:=\frac12(\alpha^2+\beta^2).\label{setti}\eeq 
In some cases the parameter $\kappa$ will be replaced by
\beq\tau:=\frac\i2(\alpha^2-\beta^2)=\i\kappa.\eeq

We will describe two complex classes of exactly solvable potentials depending
on two complex parameters. Within
each complex class there will be three
 real classes of exactly solvable potentials
depending on two real parameters.

\subsection{Trigonometric P\"oschl-Teller potential}

In this subsection we consider Ansatz (1) of (\ref{cases}).
We set \beq z=\sin^2\frac{r}{2},\ \ \hbox{ which solves }
\ z'=z^{\frac12}(1-z)^{\frac12}.\label{a1}\eeq
This leads to the Schr\"odinger equation
\beq \left(-\partial_r^2+V_{\delta,\kappa}^{\rm tPT}(r)-\frac{\mu^2}{4}\right)\phi(r)=0,
\label{posch}\eeq
where
\begin{eqnarray}\label{posch1}
V_{\delta,\kappa}^{\rm tPT}(r)&:=&
\left(\alpha^2-\frac{1}{4}\right)\frac{1}{4\sin^2\frac{r}{2}}+
\left(\beta^2-\frac{1}{4}\right)\frac{1}{4\cos^2 \frac{r}{2}}
\\&=&
\left(\delta-\frac{1}{4}\right)\frac{1}{\sin^2 r}+
\kappa\frac{\cos r}{\sin^2 r}
.\nonumber\end{eqnarray}
This potential was proposed and solved by G.~P\"oschl and E.~Teller 
\cite{pt}. It is usually called the {\em P\"oschl-Teller potential}, sometimes
also
the {\em P\"oschl-Teller potential of the first kind}
 or the {\em trigonometric
  Scarf potential}.

A natural real domain for this potential is $]0,\pi[$.
If $\kappa,\delta$ are real, then the potential is real on this domain.

Explicitly, the reduction of (\ref{posch}) to the hypergeometric equation
is derived as follows:
\begin{eqnarray}
&&-z^{\frac{\alpha}{2}+\frac{1}{4}}(1-z)^{\frac{\beta}{2}+\frac{1}{4}}
\F\left({\textstyle\frac{\alpha+\beta+\mu+1}{2}},
{\textstyle\frac{\alpha+\beta-\mu+1}{2}}
;1+\alpha;z,\p_z\right)
z^{-\frac{\alpha}{2}-\frac{1}{4}}(1-z)^{-\frac{\beta}{2}-\frac{1}{4}}
\nonumber\\[2ex]
&=&-z(1-z)\P\left(\begin{array}{cccc}
0&1&\infty&\\
\frac{\alpha}{2}+\frac{1}{4}&\frac{\beta}{2}+\frac{1}{4}&\frac{\mu}{2}&z,\p_z\\
-\frac{\alpha}{2}+\frac{1}{4}&-\frac{\beta}{2}+\frac{1}{4}&-\frac{\mu}{2}
&\end{array}\right)\nonumber\\[4ex]
&=&-z(1-z)\left(\partial_z^2+\left(\frac{1}{2z}-\frac{1}{2(1-z)}\right)
\partial_z\right)
\nonumber\\&&
+\left(\alpha^2-\frac14\right)\frac{1}{4z}
+\left(\beta^2-\frac14\right)\frac{1}{4(1-z)}-\frac{\mu^2}{4}\label{riem}
\\\nonumber
&=&-\partial_r^2+\Big(\alpha^2-\frac{1}{4}\Big)\frac{1}{4\sin^2\frac{r}{2}}+
\Big(\beta^2-\frac{1}{4}\Big)\frac{1}{4\cos^2\frac{ r}{2}}
-\frac{\mu^2}{4}
.
\end{eqnarray}

\subsection{Hyperbolic P\"oschl-Teller potential}
We continue with (1) of (\ref{cases}). We set
 \beq z=-\sinh^2\frac{r}{2},\ \ \hbox{ which solves }
\ z'=-(-z)^{\frac12}(1-z)^{\frac12}.\label{a2}\eeq
This leads to the Schr\"odinger equation
\beq \left(-\partial_r^2+V_{\delta,\kappa}^{\rm hPT}(r)+\frac{\mu^2}{4}
\right)\phi(r)=0 \label{hpt},
\label{hyper}\eeq
where\begin{eqnarray}V_{\delta,\kappa}^{\rm hPT}(r)&:=&
\left(\alpha^2-\frac{1}{4}\right)\frac{1}{4\sinh^2\frac{r}{2}}-
\left(\beta^2-\frac{1}{4}\right)\frac{1}{4\cosh^2 \frac{r}{2}}
\\&=&
\left(\delta-\frac{1}{4}\right)\frac{1}{\sinh^2 r}+
\kappa\frac{\cosh r}{\sinh^2 r}\nonumber
.\label{a2a}\end{eqnarray}
This potential was also proposed and solved by G.~P\"oschl and E.~Teller 
\cite{pt}. In the literature it is  known as {\em hyperbolic, generalized 
P\"oschl-Teller potential},
or the 
{\em P\"oschl-Teller potential of the second kind}.

A natural real domain for this potential is $]0,\infty[$.
If  $\kappa,\delta$ are real, then the potential is real on this domain.

To see that (\ref{hpt}) can be solved in terms of the hypergeometric equation,
we first repeat the computations leading to (\ref{riem}), and then set 
$z=-\sinh^2\frac{r}{2}$.

\subsection{Scarf potential}
We continue with (1) of (\ref{cases}). We set
 \beq z=\frac12-\i\cosh\frac{ r}{2}\sinh \frac{r}{2},\ \ \hbox{ which solves }
\ z'=(-z)^{\frac12}(1-z)^{\frac12}.\label{a2b}\eeq
This leads to the Schr\"odinger equation
\beq \left(-\partial_r^2+V_{\delta,\tau}^{\rm S}(r)+\frac{\mu^2}{4}\right)\phi(r)=0,
\label{hyper5}\eeq
where\begin{eqnarray}V_{\delta,\tau}^{\rm S}(r)&:=&
-\left(\delta-\frac{1}{4}\right)\frac{1}{\cosh^2 r}-
\tau\frac{\sinh r}{\cosh^2 r}\nonumber
.\label{a2c}\end{eqnarray}
This potential was proposed and solved by F. Scarf \cite{scarf}.
 In the literature it is  often called the {\em hyperbolic Scarf potential}.

A natural real domain for this potential is $]-\infty,\infty[$.
If  $\delta,\tau$ are real, then the potential is real on this domain.

To see that (\ref{hyper5}) can be solved in terms of the hypergeometric equation,
we first repeat the computations leading to (\ref{riem}), and then set 
$z=\frac12-\i\cosh \frac{r}{2}\sinh \frac{r}{2}$.

\subsection{More about  P\"oschl-Teller and Scarf potentials}

Both kinds of
P\"oschl-Teller potentials and the Scarf potential 
are real cases of the same complex  case. To see
this, consider eg. the hyperbolic P\"oschl-Teller potential
 as a function of the
complex parameter $r$. It is holomorphic away from singularities at
$\i\pi n$, $n\in\zz$. 

For real $\delta,\kappa$, $V_{\delta,\kappa}^{\rm hPT}(r)$ is real on
$\i\rr$ and $\rr+\i\pi n$. On each halfline
$]0,\infty[+\i\pi n$ and $]-\infty,0[+ \i\pi n$ we obtain
    the hyperbolic P\"oschl-Teller potential. On each interval $]\i\pi
    n,\i\pi (n+1)[$ we obtain the trigonometric 
 P\"oschl-Teller potential. 

For real $\delta,\i\kappa,$, $V_{\delta,\kappa}^{\rm hPT}(r)$ is real on 
$\rr+\i\pi(n+\frac12)$. On each of these lines we obtain the Scarf potential.

Above, we used Ansatz (1) to derive Scarf and both kinds of P\"oschl-Teller
potentials. Alternatively, one can use
Ansatzes (2) or (3). To see this it is enough to consider Ansatz (3).
In fact, we first repeat computations analogous to (\ref{riem}):
\begin{eqnarray*}
\\[2ex]
&&-z(1-z)^2\P\left(\begin{array}{cccc}
0&1&\infty&\\
\frac{\alpha}{2}+\frac{1}{4}&\frac{\beta}{2}&\frac{\mu}{2}+\frac14&z,\p_z\\
-\frac{\alpha}{2}+\frac{1}{4}&-\frac{\beta}{2}&-\frac{\mu}{2}+\frac14
&\end{array}\right)\\[4ex]
&=&-z(1-z)^2\left(\partial_z^2+\Big(\frac{1}{2z}-\frac{1}{1-z}\Big)
\partial_z\right)
\\&&
+\left(\alpha^2-\frac14\right)\frac{1-z}{4z}
+\frac{\beta^2}{4}-\left(\mu^2-\frac14\right)\frac{(1-z)}{4}.
\end{eqnarray*}

We set \beq z=\tgh^2\frac{r}{2},\ \ \hbox{ which solves }
\ z'=z^{\frac12}(1-z),\label{b1}\eeq
obtaining the hyperbolic   P\"oschl-Teller potential,
 \beq z=-\tg^2\frac{r}{2},\ \ \hbox{ which solves }
\ z'=-(-z)^{\frac12}(1-z),\label{b2}\eeq
obtaining the trigonometric   P\"oschl-Teller potential,
or
 \beq z=\ctgh^2\frac{r}{2},\ \ \hbox{ which solves }
\ z'=z^{\frac12}(1-z),\label{b3}\eeq
obtaining the  Scarf potential.

\subsection{Manning-Rosen potential}

Let us consider Ansatz (3) of (\ref{cases}).
We set \beq z=\frac{1}{1+\e^{2r}},\ \ \hbox{ which solves }
\ z'=2z(z-1).\label{w1}\eeq
This leads to the Schr\"odinger equation
\beq \left(-\partial_r^2+V_{\kappa,\mu}^{\rm MR}(r)+\delta\right)\phi(r)=0,
\label{rosen}\eeq
where
\[
V_{\kappa,\mu}^{\rm MR}(r):=-\kappa\frac{\sinh r}{\cosh
  r}-\left(\frac{\mu^2}{4}-\frac14\right) 
\frac{1}{\cosh^2
r}.\]
This potential was proposed and solved by M.~F.~Manning and N.~Rosen
 \cite{mrosen}. In the
literature it is also called the {\em Woods-Saxon 
 potential} \cite{sw} (for instance in \cite{fluegge}), also the {\em
  hyperbolic Rosen-Morse potential}.

A natural real domain for this potential is $]-\infty,\infty[$.
The potential is real if $r\in]-\infty,\infty[$ and $\kappa,\beta^2$ are real.


Here is a derivation of (\ref{rosen}) from the hypergeometric equation:
\begin{eqnarray}\nonumber
&&-4z^{1+\frac{\alpha}{2}}(1-z)^{1+\frac{\beta}{2}}
\F({\textstyle\frac{\alpha+\beta+\mu+1}{2}},
{\textstyle\frac{\alpha+\beta-\mu+1}{2}}
;1+\alpha;z,\p_z)
z^{-\frac{\alpha}{2}}(1-z)^{-\frac{\beta}{2}}
\nonumber\\[3ex]
&=&
-4z^2(z-1)^2\P\left(\begin{array}{cccc}
0&1&\infty&\\
\frac{\alpha}{2}&\frac{\beta}{2}&\mu+\12&z,\p_z\\
-\frac{\alpha}{2}&-\frac{\beta}{2}&-\mu+\12
&\end{array}\right)\nonumber
\\[3ex]&=&
-4z^2(1-z)^2\left(\p_z^2+\Big(\frac{1}{z}-\frac{1}{1-z}\Big)\partial_z\right)
+\alpha^2(1-z)+
\beta^2z-(\mu^2-1)z(1-z)\label{riem3}
\\[3ex]&=&
-\partial_r^2
+\delta+\kappa\frac{\e^{2r}-1}{1+\e^{2r}}
-(\mu^2-1)\frac{\e^{2r}}{(1+\e^{2r})^2}.
\end{eqnarray}

\subsection{Eckart potential}
We still consider Ansatz (3) of (\ref{cases}).

We set \beq z=\frac{1}{1-\e^{-2r}},\ \ \hbox{ which solves }
\ z'=2z(1-z),\label{w2}\eeq
and use (\ref{setti}).
This leads to the Schr\"odinger equation
\beq \left(-\partial_r^2+V_{\kappa,\mu}^{\rm E}(r)+\delta\right)\phi(r)=0,\eeq
where
\[
V_{\kappa,\mu}^{\rm E}(y):=-\kappa\frac{\cosh r}{\sinh
  r}+\left(\frac{\mu^2}{4}-\frac14\right) 
\frac{1}{\sinh^2
r}.\]
This potential was proposed and solved by C.~Eckart \cite{eckart}. In the
literature (for instance \cite{fluegge}) it is also called the {\em Hulthen
  potential}  \cite{hulthen}, sometimes
also the {\em generalized Morse potential}, because of its similarity to the
Morse potential, see Subsect. \ref{sec-morse}.

A natural real domain for this potential is $]0,\infty[$.
If $\kappa,\beta^2$ are real, then the potential is real on this domain.

To derive the Eckart potential, we first repeat the computations leading to
(\ref{riem3}), and then set 
$z=\frac{1}{1-\e^{-2r}}$.

\subsection{Rosen-Morse potential}

Once again, we consider Ansatz (3) of (\ref{cases}).

We set \beq z=\frac{1}{1-\e^{2\i r}},\ \ \hbox{ which solves }
\ z'=2\i z(1-z).\label{w2a}\eeq
This leads to the Schr\"odinger equation
\beq \left(-\partial_r^2+V_{\tau,\mu}^{\rm RM}(r)-\delta\right)\phi(r)=0,\eeq
where
\[
V_{\tau,\mu}^{\rm RM}(y):=\tau\frac{\cos r}{\sin  r}+
\left(\frac{\mu^2}{4}-\frac14\right) 
\frac{1}{\sin^2
r}.\]
This potential is known as the {\em Rosen-Morse potential}, also the
{\em trigonometric Rosen-Morse potential} (altough this name is widely used in the literature, we were unable to explain decisively its origin).

A natural real domain for this potential is $]0,\pi[$. If $\tau,\mu^2$ are real, then the potential is real on this domain.

To derive the Rosen-Morse potential, we 
 first repeat the computations leading to
(\ref{riem3}), then set 
$z=\frac{1}{1-\e^{2\i r}}$.
\subsection{More about
Manning-Rosen, Eckart and  Rosen-Morse potentials} 

The Manning-Rosen, Eckart and  Rosen-Morse potentials are all real cases of
the same complex case. 
To see
this, consider eg. the Eckart potential as a function of the
complex parameter $r$. It is holomorphic away from singularities at
$\i\pi n$, $n\in\zz$. 

For real $\kappa,\mu^2$, $V_{\kappa,\mu}^{\rm E}(r)$ is real on
the lines $\rr+\frac{\i\pi n}{2}$. On each line $\rr+\i (m+\frac12)
\pi$, $m\in\zz$,
 we obtain the Manning-Rosen
 potential. On each halfline
$]0,\infty[+ \i\pi m$ and $]-\infty,0[+ \i\pi m$,  we obtain
 the    Eckart potential. 

For real $\i\kappa,\mu^2$,  $V_{\kappa,\mu}^{\rm E}(r)$ is real on
$\i\rr$. On each interval $]\i\pi m,\i\pi(m+1)[$ we obtain the
    Rosen-Morse potential.

Ansatzes (5) and (6) lead to the same classes of exactly solvable
potentials as Ansatz (4). To see this it is enough to consider Ansatz (5).
In fact, we first repeat computations analogous to (\ref{riem3}):

\begin{eqnarray*}
&&-4z^{-\frac{\alpha+1}{2}}(1-z)^{2+\frac{\beta}{2}}
\F({\textstyle\frac{\alpha+\beta+\mu+1}{2}},
{\textstyle\frac{\alpha+\beta+-\mu+1}{2}};
1+\alpha;z,\p_z)
z^{-\frac{\alpha+1}{2}}(1-z)^{-\frac{\beta}{2}}
\\[3ex]&=&
-4(1-z)^2\P\left(\begin{array}{cccc}
0&1&\infty&\\
\alpha+\12&\frac{\beta}{2}&\frac{\mu}{2}&z,\p_z\\
-\alpha+\12&-\frac{\beta}{2}&-\frac{\mu}{2}
&\end{array}\right)\\[3ex]
\\[2mm]&=&
-4(1-z)^2\left(\p_z^2-\frac{1}{1-z}\partial_z\right)\\&&+
(\alpha^2-1)\frac{1-z}{z^2}+
\beta^2\frac{1}{z}-\mu^2\frac{1-z}{z}.
\end{eqnarray*}
We set \beq z=1-\e^{2r},\ \ \hbox{ which solves }
\ z'=-2(1-z),\label{d1}\eeq
obtaining the Eckart potential,
 \beq z=1+\e^{2r},\ \ \hbox{ which solves }
\ z'=2(1-z),\label{d2}\eeq
obtaining the Manning-Rosen potential,
 \beq z=1-\e^{\i2r},\ \ \hbox{ which solves }
\ z'=-2\i(1-z),\label{d3}\eeq
obtaining the Rosen-Morse potential,

\section{Schr\"odinger operators reducible to the rescaled confluent equation}

One of basic exactly solvable equations is the 
{\em confluent
equation}, given by the operator
\beq \cF(a;c;z,\partial_z):=z\p_z^2+(c- z)\p_z-a.\label{conf1}\eeq
It is convenient to consider (\ref{conf1}) in parallel with
the equation given by the operator
\beq \F(a,b;-;z,\p_z):=z^2\p_z^2+(-1+(1+a+b)z)\p_z+ab.
\label{g8}\eeq
The equation given by (\ref{g8}) is sometimes called the  {\em ${}_2F_0$ equation}.
Note that
\begin{eqnarray*}
z^{a}\F(a,b;-;z,\p_z)z^{-a}
&=&w\F(a;1+a-b;w,\p_w),\ \ w=-z^{-1},\ \ z=-w^{-1}.
\label{g8a}\end{eqnarray*}
Hence the ${}_2F_0$ equation is equivalent to the confluent
equation.
The relationship between the parameters is
\[c=1+a-b,\ \ \ \ b=1+a-c.\]

Another exactly solvable 
equation that we will consider in this section is sometimes called
the {\em ${}_0F_1$ equation}. It is given by
\beq \cF(c;z,\partial_z):=z\p_z^2+(c- z)\p_z-a,\eeq
and is equivalent to the {\em Bessel equation}.

Clearly, the confluent, ${}_2F_0$ and ${}_0F_1$ equations belong to the class of
hypergeometric type equations.

The basic equation of this section will be (\ref{hyp2}), given by
\beq
\F(a;c;\gamma;z;\p_z):=z\p_z^2+(c-\gamma z)\p_z-a.\label{conflu}\eeq
It will be called the {\em rescaled confluent equation}.
Note that in the case $\gamma=1$ (\ref{conflu}) coincides with the confluent
equation. 
If $\gamma\neq0$,  (\ref{conflu}) can be reduced to the confluent equation by
rescaling (and hence also to the ${}_2F_0$ equation).
 If $\gamma=0$,  (\ref{conflu}) coincides with the
 ${}_0F_1$ equation.

The Bose invariant of the rescaled confluent equation equals
\begin{eqnarray*}
{}_1I_1(a,b;\gamma;z)
&=&\frac{-\gamma^2z^2+2(c\gamma-2a)z+2c-c^2}{4z^2}
.
\end{eqnarray*}

\subsection{Symmetries of the rescaled confluent equation}

Let us first describe symmetries of the rescaled confluent equation. 
\begin{eqnarray}
z^{c-1}\F(a;c;\gamma;z,\p_z)z^{1-c}&=&\F(a-c\gamma+\gamma;2-c;\gamma;z,\p_z);
\label{po1}\\
\e^{-\gamma z}
\F(a;c;\gamma;z,\p_z)
\e^{\gamma z}&=&-\F(c-a;c;\gamma;w,\p_w),\ \ \ z=-w;
\label{po2}\end{eqnarray}
Besides, the scaling acts as follows:
\[\F(a;c;\gamma;z,\p_z)=\gamma\F(a/\gamma;c;1;w,\p_w), \ \ w=\gamma z.\]

It is convenient to  introduce new parameters
$\alpha,\nu $: 
\[\begin{array}{ll}
a=\frac{\gamma+\alpha\gamma-\nu }{2},& 
 c=1+\alpha;\\[2ex]
\alpha=c-1=a-b,& \nu =c\gamma-2a=1-a-b.\end{array}
\]
In the new parameters the Bose invariant of the rescaled confluent equation
has a more symmetric form:
\begin{eqnarray*}
{}_1I_1(\alpha,\nu,\gamma;z)
&=&-\frac{\gamma^2}{4}+\frac{\nu}{2 z}+\frac{1-\alpha^2}{4z^2}.
\end{eqnarray*}
Thus the starting point for the further analysis will be the equation
\beq
\left(-\partial_z^2+\frac{\gamma^2}{4}-\frac{\nu}{2z}+\Big(\frac{\alpha^2}{4}-\frac{1}{4}\Big)\frac{1}{z^2}\right)\phi(z)=0.\label{start}\eeq

We will describe 3 classes of Schr\"odinger operators solved
using the confluent equation  corresponding to three obvious choices for
$\Big(r'(z)\Big)^2$: 
\beq (1)\ \frac{1}{4},\ \ (2)\ \frac{1}{4z},\ \ (3)\ \frac{1}{4z^2}.
\label{cases1}\eeq

\subsection{Hydrogen atom}
We consider (1) of (\ref{cases1}).
We set
 \beq z=2r.\label{ac2}\eeq
This leads to the Schr\"odinger equation
\beq \left(-\partial_r^2+V_{\alpha,\nu}(r)+\gamma^2\right)\phi(r)=0,
\label{hyperc}\eeq
with the potential
\[
V_{\alpha,\nu}(r):=-\frac{\nu}{r}+
\left(\frac{\alpha^2}{4}-\frac{1}{4}\right)\frac{1}{r^2}. 
\]
(\ref{hyperc}) is the radial part of the Schr\"odinger equation for the
hydrogen atom.

$]0,\infty[$ is a natural real domain for this equation. If $\nu,\alpha^2$ are
    real, then so is $V_{\alpha,\nu}$ on $]0,\infty[$.

The derivation of (\ref{hyperc}) from (\ref{start}) is immediate:
\begin{eqnarray*}
&&4\left(-\partial_z^2+\frac{\gamma^2}{4}-\frac{\nu}{2z}+
\Big(\frac{\alpha^2}{4}-\frac{1}{4}\Big)\frac{1}{z^2}\right) \\
&=&
-\partial_r^2+\gamma^2-\frac{\nu}{r}
+\Big(\frac{\alpha^2}{4}-\frac{1}{4}\Big)\frac{1}{r^2}.
\end{eqnarray*}


\subsection{Rotationally symmetric  harmonic oscillator}

We consider (2) of (\ref{cases1}). We set 
\beq z=r^2, \ \ \ \hbox{which solves }\ \ \  z'=2\sqrt z.\eeq

This leads to the Schr\"odinger equation
\beq \left(-\partial_r^2+V_{\alpha,\gamma}(r)-2\nu\right)\phi(r)=0,
\label{hyperc1}\eeq
with the potential
\[
V_{\alpha,\gamma}(r):=\gamma^2r^2+\left(\alpha^2-\frac{1}{4}\right)
\frac{1}{r^2}.\]

$]0,\infty[$ is a natural real domain. For real $\gamma^2,\alpha^2$, the
    potential is real on $]0,\infty[$.
 By scaling, the only different real cases are
    $\gamma^2=1,0,-1$.

For $\gamma^2=1$ the equation is the radial part of the rotationally symmetric 
harmonic oscillator. For $\gamma^2=0$ the equation is the radial part of the
Helmholtz equation.


 Here is an explicit derivation of (\ref{hyperc1}) from
(\ref{start}):
\begin{eqnarray*}
&&4z^{1-\frac14}\left(-\partial_z^2+\frac{\gamma^2}{4}-\frac{\nu}{2z}+
\Big(\frac{\alpha^2}{4}-\frac{1}{4}\Big)\frac{1}{z^2}\right)z^{\frac14} \\
&=&-4z\partial_z^2-2\partial_z+\gamma^2z-2\nu+
\Big(\alpha^2-\frac{1}{4}\Big)\frac{1}{z}\\
&=&-\partial_r^2+\gamma^2r^2-2\nu+
\Big(\alpha^2-\frac{1}{4}\Big)\frac{1}{r^2}.\end{eqnarray*}

\subsection{Morse potential}
\label{sec-morse}

We consider (3) of (\ref{cases1}).
 We set 
\beq z=\e^{-r}, \ \ \ \hbox{which solves }\ \ \  z'= -z.\eeq

This leads to the Schr\"odinger equation
\beq \left(-\partial_r^2+V_{\nu,\gamma}(r)+\frac{\alpha^2}{4}\right)\phi(r)=0,
\label{hyperc2}\eeq
where
\[
V_{\nu,\gamma}(r)=\frac{\gamma^2}{4}\e^{-2r}-\frac{\nu}{2}\e^{-r}.\]

$]-\infty,\infty[$ is a natural real domain. For real $\nu,\gamma^2$, the
    potential is real on $]-\infty,\infty[$.
 By translation, the only different real cases are
    $\gamma^2=1,0,-1$.

 Here is an explicit derivation of (\ref{hyperc2}) from
(\ref{start}):
\begin{eqnarray*}
&&z^{2-\frac12}\left(-\partial_z^2+\frac{\gamma^2}{4}-\frac{\nu}{2z}+
\Big(\frac{\alpha^2}{4}-\frac{1}{4}\Big)\frac{1}{z^2}\right)z^{\frac12} \\
&=&-z^2\partial_z^2-z\partial_z+\frac{\gamma^2}{4}z^2-\frac{\nu}{2}z+
\frac{\alpha^2}{4}\\
&=&-\partial_r^2+\frac{\gamma^2}{4}\e^{-2r}-\frac{\nu}{2}\e^{-r}+
\frac{\alpha^2}{4}.\end{eqnarray*}

\section{Schr\"odinger equations reducible to the translated harmonic
  oscillator} 

The last family of 
exactly solvable Schr\"odinger equations that we describe is (\ref{harmo}),
given by
\beq 
  -\partial_z^2+\theta^2 z^2+\rho z+\lambda.\label{harmo1}\eeq
We will call (\ref{harmo1}) the {\em translated harmonic oscillator
  equation}. It is already in the canonical form.

If  $\theta^2\neq0$, then  (\ref{harmo1}) is  just the translation
of the Schr\"odinger equation for
usual
harmonic oscillator. It is then equivalent to the Hermite equation, given by
\[ \cG(a,y,\partial_y):=(\partial_y^2-2y\partial_y-2a).\]
In fact, we have
\begin{eqnarray*}
&&\theta\e^{-\frac{y^2}{2}}\cG(a,y,\partial_y)\e^{\frac{y^2}{2}}\\
&=&\theta\left(-\partial_y^2+y^2+2a-1\right)\\
&=&-\partial_z^2+\theta^2z^2+\rho z +\frac{\rho^2}{2\theta^2}+2a-1,
\end{eqnarray*}
where $y=\sqrt\theta\left(z+\frac{\rho}{2\theta^2}\right)$.

For $\theta=0$, $\rho\neq0$, (\ref{harmo1}) is equivalent to the Airy equation
\[(\partial_y^2+y)\psi(y)=0,\]
which in turn is equivalent to a special case of the ${}_0F_1$ equation:
\begin{eqnarray*}
&&-\partial_z^2+\rho
z+\lambda\\
&=&\rho^{\frac23}\left(-\partial_y^2+y\right)\\
&=&
(\rho3)^{\frac23}w^{\frac13}\F\left(\frac23;w,\p_w\right),\end{eqnarray*}
where we set $y=\rho^{\frac13} z+\rho^{-\frac23}\lambda$, and 
$w=3^{-2}y^3$.

We will describe 3 classes of Schr\"odinger operators that can be reduced to
(\ref{harmo1})   corresponding to three obvious choices for
$\Big(r'(z)\Big)^2$: 
\beq (1)\ 1,\ \ \ \ \ (2)\ z,\ \ \ \ \ (3)\ z^2.
\label{cases2}\eeq

\subsection{Translated harmonic oscillator}

We consider Ansatz (1) of (\ref{cases2}), which
corresponds to 
the most obvious choice of the energy, that is $-\lambda$. We rename the
variable $z=r$.
This leads to the Schr\"odinger equation
\beq \left(-\partial_r^2+V_{\theta,\rho}(r)+\lambda\right)\phi(r)=0,
\label{harmo4}\eeq
where
\begin{eqnarray*}
V_{\theta,\rho}(r)=
\theta^2 r^2+\rho
r.\end{eqnarray*}

\subsection{Special potential I}
\label{specI}

We can choose the energy to be $-\rho$, that is Ansatz (2) of (\ref{cases2}).
This corresponds to the substitution
\[z=\left(\frac{3r}{2}\right)^{\frac23},\ \ \hbox{which solves}
\ \ \ z'=z^{-\frac12},\] and
 leads to the Schr\"odinger equation
\beq \left(-\partial_r^2+V_{\theta,\lambda}(r)+\rho\right)\phi(r)=0,
\label{harmo2}\eeq
where
\begin{eqnarray*}
V_{\theta,\lambda}(r)=
\theta^2\left(\frac{3r}{2}\right)^{\frac23}+
\lambda\left(\frac{2}{3r}\right)^{\frac23}-\frac{5}{36}\frac{1}{r^2}
.\end{eqnarray*}
In fact,
\begin{eqnarray*}&&
z^{\frac14-1}\left(-\partial_z^2+\theta^2 z^2+\rho
z+\lambda\right)z^{-\frac14}\\
&=&-\frac1z\partial_z^2+\frac{1}{2z^2}\partial_z+\theta^2
z+\rho+\frac{\lambda}{z}
-\left(\frac{1}{4}+\frac{1}{4^2}\right)\frac{1}{z^3}\\
&=&-\partial_r^2+\theta^2\left(\frac{3r}{2}\right)^{\frac23}+\rho+
\lambda\left(\frac{2}{3r}\right)^{\frac23}-\left(\frac{1}{4}+\frac{1}{4^2}\right)
\left(\frac{2}{3r}\right)^2. \end{eqnarray*}

Note the following intriguing feature of the above potential: the coefficient
at $r^{-2}$ is fixed and one cannot change it by rescaling the variable $r$.
\subsection{Special potential II}
\label{specII}
We can choose the energy to be $-\theta^2$,
 that is Ansatz (3) of (\ref{cases2}). 
This corresponds to the substitution
\[z=(2r)^{\frac12},\ \ \hbox{which solves}\ \ \ 
z'=z^{-1},\] and
 leads to the Schr\"odinger equation
\beq \left(-\partial_r^2+V_{\rho,\lambda}(r)+\theta^2\right)\phi(r)=0,
\label{harmo3}\eeq
where
\begin{eqnarray*}
V_{\rho,\lambda}(r)=
\frac{\rho}{(2r)^{\frac12}}+
\frac{\lambda}{2r}-\frac{3}{16}\frac{1}{r^2}
.\end{eqnarray*}
In fact,
\begin{eqnarray*}&&
z^{\frac12-2}\left(-\partial_z^2+\theta^2 z^2+\rho
z+\lambda\right)z^{-\frac12}\\
&=&-\frac{1}{z^2}\partial_z^2+\frac{1}{z^3}\partial_z+\theta^2
+\frac{\rho}{z}+\frac{\lambda}{z^2}
-\left(\frac{1}{2}+\frac{1}{2^2}\right)\frac{1}{z^4}\\
&=&-\partial_r^2+\theta^2+\frac{\rho}{(2r)^{\frac12}}+\frac{\lambda}{2r}-
\left(\frac{1}{2}+\frac{1}{2^2}\right)\frac{1}{2^2r^2}
. \end{eqnarray*}

Again, the coefficient
at $r^{-2}$ is fixed and one cannot change it by rescaling the variable $r$.

\end{document}